# Ballisticity of nanotube FETs: Role of phonon energy and gate bias

Siyuranga O. Koswatta , Sayed Hasan, and Mark S. Lundstrom

School of Electrical and Computer Engineering, Purdue University, West Lafayette,

Indiana 47907-1285

M. P. Anantram

Center for Nanotechnology, NASA Ames Research Center, MS 229-1, Moffett Field,

California 94035-1000

Dmitri E. Nikonov[b]

Technology and Manufacturing Group, Intel Corp., SC1-05, Santa Clara, California

95052

We investigate the role of electron-phonon scattering and gate bias in degrading the drive current of nanotube MOSFETs. Our central results are: (i) Optical phonon scattering significantly decreases the drive current only when gate voltage is higher than a well-defined threshold. It means that elastic scattering mechanisms are most detrimental to nanotube MOSFETs. (ii) For comparable mean free paths, a lower phonon energy leads to a larger degradation of drive current. Thus for semiconducting nanowire FETs, the drive current will be more sensitive than carbon nanotube FETs because of the smaller phonon energies in semiconductors. (iii) Radial breathing mode phonons cause an appreciable reduction in drive current.

---

[b] Electronic address: dmitri.e.nikonov@intel.com

Interconnects and field-effect transistors (FETs) based on metallic and semiconducting nanotubes respectively, have received considerable attention over the last five years because their performance metrics make them promising electronic devices. Experiments have established that transport in metallic carbon nanotube wires is close to the ballistic limit at biases smaller than approximately 100 meV.[1, 2] At larger biases, the current carrying capacity of metallic nanotubes is far from ballistic due to zone boundary and optical phonon scattering.[3,4,5,6] In the area of nanotube FETs, experiments have demonstrated near-ballistic transport[7] and on-current performance that exceeds that of silicon transistors. [7],[8] Our objective in this letter is to provide insight into how various electron-phonon scattering mechanisms and the applied gate bias affect the on-current of a carbon nanotube MOSFET. Recent work showed that elastic scattering can strongly degrade the on-current but strong optical phonon scattering may have little effect.[9] This work extends prior work[10,11,12,13] by examining the phonon modes most likely to couple to electrons and by exploring the gate bias dependence of electron-phonon scattering. We show that the radial breathing mode may play an important role in some experiments and the gate bias dependence displays a clear threshold effect.

The current-voltage characteristics are calculated using the non-equilibrium Green's function approach,[14] where the charge density is calculated using the efficient algorithm of Ref. 15. The NEGF transport equations are solved in mode-space using the $\pi$ orbital tight binding Hamiltonian, self-consistently with Poisson's equation.[16] Electron-phonon scattering is treated within the mode space approach in a manner similar to [9,16], but we carefully consider both the longitudinal (L) and transverse (T) optical (OP) and acoustic (AP) phonon branches, and the radial breathing modes (RBM).[17] In this letter we disregard the real part of the self energy and focus on the imaginary part which corresponds to scattering.

The CNT MOSFET structure employed in this study, shown in Figure 1, is a (13,0) zigzag CNT with bandgap, $E_G$ = 0.8eV, channel length, $L_G$ = 20nm, and high-k HfO$_2$ cylindrical gate with thickness $t_{ox}$ = 2nm. The lowest electronic sub-bands in the (13,0) nanotube correspond to two degenerate valleys with azimuthal quantum numbers $m$ = 9 and $m$ = 17. Therefore the phonons with azimuthal quantum numbers of $m$ = 0 cause intra-valley transitions and phonons with $m$ = 8 cause the inter-valley transitions (also know as zone-boundary scattering). For moderate biases electrons are concentrated close to the band minima, and we can consider phonons with $q \approx 0$ (long wavelength). The optical and RBM phonon modes considered in this study are calculated according to [17] and the corresponding electron-phonon coupling constants (EPC) are summarized in Table 1. The electron-phonon coupling strengths for these modes in an (n,0) zigzag CNT are determined by

$$R_{e-OP} = \frac{J_1^2 \hbar |M_{OP}|^2}{n M_C \omega_{OP}} \quad (1)$$

where $M_C$ the mass of a carbon atom, the deformation potential, $J_1$ = 6eV/Å, $\omega_{OP}$ is the frequency of the phonon mode, and the reduced matrix elements $|M_{OP}|^2$ are calculated according to [18].

For the case of acoustic phonons (with energy close to zero for the considered phonon momenta) the electron-phonon coupling strength is determined by the following phenomenological equation [7],

$$R_{e-AP} = \frac{3 a_{cc} J_0^2}{\lambda_{AP}} \quad (2)$$

where $\lambda_{AP}$ is the nominal mean free path (mfp) for acoustic phonon scattering, the hopping parameter $J_0 = 3eV$,[18] and the carbon-carbon bond length, $a_{cc} = 1.42$Å. Note that the nominal mfp is valid only for the linear portion of the electronic dispersion curve at large energies. The effective mfp, and the corresponding scattering rate, are different for the electrons close to the band minimum and are thus affected by the density of states. The energy dependent scattering rate, $\tau_{AP}(E)$, is given by,

$$\frac{1}{\tau_{AP}(E)} = \frac{2}{\hbar}\left|\text{Im}\left(\Sigma^{scat}(E)\right)\right| = \frac{2}{\hbar}R_{e-AP}\left(2\pi LDOS(E)\right) \qquad (3)$$

where $\Sigma^{scat}(E)$ is the self-energy function for electron-phonon interaction, and *LDOS(E)* is the local density of states at a given channel position.[14]

We first consider the role of scattering in influencing the $I_{DS}$-$V_{DS}$ characteristics for various phonon modes with non zero energy (Fig. 2). We find that the reduction of drain current due to scattering increases with gate bias. This bias dependence is better understood by examining the ballisticity (defined as the ratio between current under scattering and the ballistic current, $I_{scat}/I_{ballist}$) as shown in Figure 3. Ballisticity is plotted against $\eta_{FS}$, the energy difference between the source Fermi-level and the channel injection barrier. (The significance of $\eta_{FS}$, is explained by the inset of Figure 3.) Under ballistic operation, at a fairly large drain bias, only the positive going states in the channel region are filled by the source Fermi distribution. In order for electrons to back-scatter due to phonon emission, two main requirements must be satisfied: 1) the positive going electrons should be higher than the band edge by at least the amount of phonon energy, 2) there should be an empty negative going state that the electron can back-scatter into. Note that it is the scattering in the channel that matters most. When electrons reach the drain region they can emit optical phonons and scatter into empty low-energy states. However, *if the phonon energy is large*, such back-scattered electrons would not have enough energy to cross over the channel injection barrier and reach the source region, and the scattering process has a minimal effect on steady-state current transport.[9]

Under low $V_{GS}$, the conduction band edge in the channel is above $E_{FS}$, and $\eta_{FS}<0$. The electronic states in the channel with much higher energies than the source Fermi energy ($E_{FS}$) are occupied with an exponentially decaying tail of the Fermi distribution. Thus only a small fraction of the electrons crossing the barrier has enough energy to experience scattering by an optical phonon. This corroborates the results [9] that ballisticity for optical phonons can be close to unity. Note in figure 3 that for negative $\eta_{FS}$, lower energy phonons, especially with energies comparable to thermal energy, $k_B T$, (e.g.: 28meV intra-valley RBM), cause a larger decrease of current, since they can scatter back to the source overcoming the channel injection barrier.[19] The importance of RBM scattering in CNTs for carrier transport has recently been reported.[20,21,22] Experimentally, the effect of various phonon modes on current transport can be manifested in comparison of suspended and covered nanotube transistors.[23] In the former, all phonon modes are present. In the latter, the out-of plane phonon modes (RBM) are likely to be effectively clamped. Thus we expect transport to be more ballistic, and current to be higher in transistors with covered nanotubes. The presence or absence of the RBM may explain recent experiments.[23]

On the other hand, for higher $V_{GS}$ when $\eta_{FS}>0$, a significant fraction of electrons crossing the channel injection barrier is back-scattered by every phonon mode present. As a result, the ballisticity decreases significantly when $\eta_{FS} \geq \hbar\omega$. Examining figure 3 confirms that the onset of ballisticity roll-off for progressively higher phonon energies occurs at increasingly larger gate biases. As stated in Table I, the higher energy phonons (e.g.: inter-valley LO/TA (183meV) and intra-valley LO (195meV)) have greater electron-phonon coupling constants. This results in a greater decrease of current (i.e. lower ballisticity) due to scattering off these phonon branches at larger $V_{GS}$, once such scattering becomes effective.

Figure 3 also shows the ballisticity when all phonon modes of Table I are simultaneously active ("all-modes"). In most of the cases, the decrease of current is a combination of decrease from each mode. The interesting exception is that ballisticity for "all-modes" is higher than for the 28meV RBM alone at low gate bias. In the case of low-energy phonons, such as the 28meV RBM mode, back-scattering near the drain region causes carriers with enough energy to cross the channel barrier and to reach the source region, thereby contributing to reduction of ballisticity. But, in the presence of high-energy phonons, carriers near the drain are scattered into lower energy states that cannot overcome the channel barrier, thus enhancing the overall ballisticity.

Finally, we discuss ballisticity under acoustic phonon scattering. For acoustic phonon scattering, the nominal mfps ($\lambda_{AP}$) investigated in this study are as listed in Figure 4: 1500nm down to 200nm. Experimentally deduced mfps for metallic CNTs are reported to be anywhere from 300nm to more than 1μm.[3,7,8] It is interesting that the overall change in ballisticity with elastic scattering (Fig. 4) is very different from the inelastic scattering case (Figs. 3). The ballisticity with only acoustic (elastic) phonon scattering increases by about 5% with gate bias while inelastic scattering causes the ballisticity to decrease, with gate bias. The reason for this is at low $V_{GS}$ (off-state), the current is primarily determined by carriers at the conduction band edge at the channel injection barrier. Due to the large density of states near the conduction band-edge in 1D systems (van Hove singularity), such carriers experience a higher scattering rate as indicated by equation 3. Therefore, they have a smaller effective mfp compared to the channel length, resulting in a reduction in the ballisticity. However, at large $V_{GS}$, significant fraction of carriers occupy states well-above the conduction band-edge where scattering rate is much lower, resulting in an increase in ballisticity. It is quite surprising that the ballisticity is as small as 0.75 when the nominal mean free path (evaluated for the linear energy dispersion) is about ten times larger than the channel length.

In this Letter, we discussed the physics of drive current degradation (ballisticity) of nanotube MOSFETs due to electron-phonon scattering. The influence of phonon scattering on drive current is found to depend significantly on gate bias. A general conclusion applicable to all quasi one-dimensional transistors is that scattering by energetic phonons reduces the drive current only when the gate bias exceeds a threshold voltage that moves the Fermi level in the source higher than the band-edge in the channel by approximately the energy of a phonon. If only high-energy phonons are present, scattering by phonon emission may occur near the drain, but the drive current will be unaffected is the gate voltage is below the threshold value. We also find that elastic scattering is highly detrimental to the drain current at all gate biases because carriers backscattered anywhere in the nanotube can return to the source. For typical deformation potentials used in carbon nanotubes, we find that a typical FET with a channel length of 20 nm has a ballisticity of 0.97 and 0.8 at low and high gate biases, respectively. The ballisticity at low gate biases is determined by the 28meV RBM phonon and acoustic phonon scattering. The high gate bias case is determined by all the phonon modes available, but will be dominated by the modes with the strongest electron-phonon coupling. Nevertheless, depending on the device geometry, certain phonon modes, such as the radial breathing mode, could be effectively suppressed, and higher ballisticities could be achieved.

We acknowledge the support of this work by the NSF Network for Computational Nanotechnology (NCN), the NASA Institute for Nanoelectronics and Computing (NASA INAC NCC 2-1363), NASA contract NAS2-03144 to UARC, and Intel Corporation.

**List of Tables**

| Table I<br>Phonon modes and corresponding coupling parameters<br>used in this study ||| 
|---|---|---|
| **Phonon mode** | **Energy (meV)** | **Coupling strength ($eV^2$)** |
| Intra-valley LO | 195 | $23.25 \times 10^{-3}$ |
| Intra-valley RBM | 28 | $0.95 \times 10^{-3}$ |
| Inter-valley TA/LO * | 156 | $0.19 \times 10^{-3}$ |
| Inter-valley LO/TA * | 183 | $44.64 \times 10^{-3}$ |

* Zone-boundary phonon modes that mediate inter-valley scattering are found to be a mixture of fundamental polarizations.

**List of Figure Captions**

FIG. 1. CNT-MOSFET geometry employed in this study. Cylindrical high-k gate ($HfO_2$ – k=16) and $t_{OX}$ = 2nm. (13,0) Zigzag CNT with 20nm channel length ($L_G$ = 20nm) and band-gap, $E_G$ = 0.8eV. Source/Drain doping is 1.5 dopant atoms/nm (for comparison, (13,0) CNT has 122 carbon atoms/nm).

FIG. 2. (Color online) $I_{DS}$-$V_{DS}$ under ballistic transport and phonon scattering for $V_{GS}$ = 0.4, 0.5, and 0.6V. Curves are grouped by $V_{GS}$ for clarity. The inter-TA/LO mode (not shown above) does not affect the ballisticity in the above voltage bias range.

FIG. 3. (Color online) Ballisticity ($I_{scat}/I_{ballist}$) vs. $\eta_{FS}$ under phonon scattering at $V_{DS}$ = 0.4V. Here, $\eta_{FS}$ is the energy difference between the source Fermi-level and the channel injection barrier. Negative (positive) $\eta_{FS}$ corresponds to small (large) $V_{GS}$ values. $V_{GS}$ is varied from 0.2V to 0.8V in steps of 0.05V. The inter-TA/LO mode (not shown above) does not affect the ballisticity in the above voltage bias range. Inset: at large $V_{GS}$ ballisticity decreases due to OP emission.

FIG. 4. (Color online) Ballisticity (variables same as in Figure 3) under acoustic phonon scattering for various mean free paths.

**List of Figures**

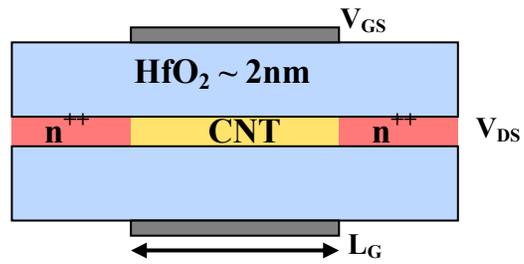

FIG. 1.

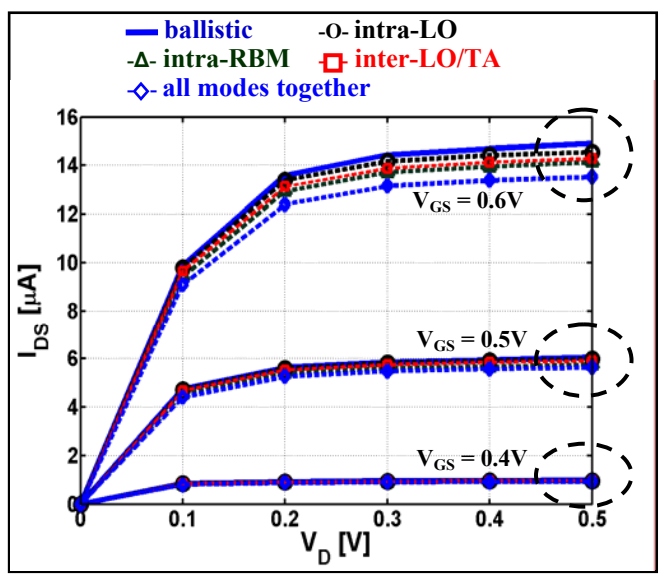

FIG. 2.

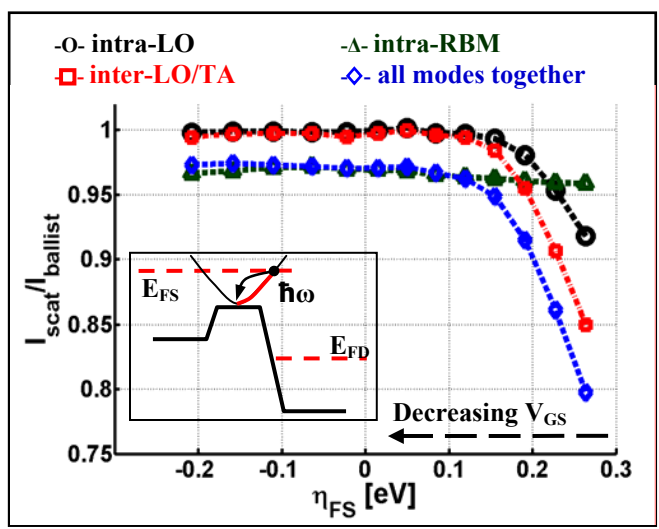

FIG. 3.

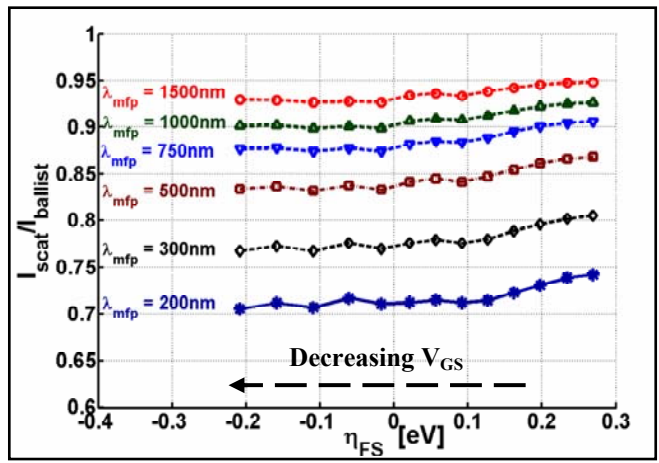

FIG. 4.